# Efficient coverage planning for full-area C-ITS communications based on radio propagation simulation and measurement tools

**Hagen Ußler[1*], Christian Setzefand[2], Daniel Kanis[2], Paul Schwarzbach[1], Oliver Michler[1]**
1. Technische Universität Dresden, Institute of Traffic Telematics, 01062 Dresden, Germany
E-Mail: {hagen.ussler, paul.schwarzbach, oliver.michler}@tu-dresden.de
2. MRK Media AG, 01067 Dresden, Germany
E-Mail: {christian.setzefand, daniel.kanis}@mrk-media.de

**Abstract:**

Intelligent infrastructure, currently often consisting of C-ITS stations and prospectively supplemented by 5G, is a key-enabler for application-oriented and area-wide realization of highly automated and connected driving. For this, radio coverage along the routes must be ensured, leading to high demands on location- and radio-specific planning and parameterization of roadside units (RSU). Hence, this paper presents efficient planning, measurement and evaluation methods for RSU coverage outlining, allowing economically efficient and technically secured planning of intelligent infrastructure. Necessary scientific technical steps are showcased along a 3.5 km testbed for automated and connected driving in rural environments. First, a radio propagation simulation based on a 3D environment model and its electro-magnetic properties is performed, allowing the examination and optimization of RSU quantity as well as site and antenna selection. Additionally, the necessary calibration of simulation results based on continuous wave (CW) and C-ITS service measurements in both lab-based and real-world scenarios is presented.

## 1. Introduction

The deployment and testing of Cooperative Intelligent Transport Systems and Services (C-ITS) has been emerging for several years. Starting with the first ITS G5 communication pilots based on the radio standard IEEE 802.11p, the recent targeted and widespread expansion of the fifth-generation (5G) mobile communications standard drives the commissioning of C-ITS infrastructure. Connected and intelligent infrastructure and vehicles are a key-enabler for automated and autonomous driving functions and significantly contribute to maintain efficient, sustainable and resilient transport routes as well as passenger and freight transport. Therefore, vehicle manufacturers and transport infrastructure operators are increasingly relying on C-ITS, whether RSU-based or 5G-driven.

The necessary digital information transmission for communication services can be realized by intelligent infrastructure components. In order to ensure a continuous supply of services, high demands on a seamless and stable radio link, especially for safety-critical applications, must be met. Several C-ITS stations are required along roadways to ensure radio coverage of large-scale traffic networks.

The planning of conventional communication infrastructures along roads is often lengthy, cost-intensive and subject to strict regulations and requirements. Due to the importance of digital road infrastructure, C-ITS components quantity, locations and antenna systems must be conscientiously planned and implemented under radio-technological conditions without a trial-and-error procedure. This helps to counter the trade-off between inadequate radio coverage or high costs due to over-deployment.

Software-based radio planning tools can be used to plan resilient, high-performance future road infrastructures. These can be used to simulate, visualize, and evaluate both empirical (e.g., COST 231) and deterministic (e.g., 3D ray-tracing or dominant path) radio signal propagation simulations within CAD-based environmental models.

The advantage of radio simulation is the relatively accurate calculation of radio coverage in specific planning environments, which provides a fast and efficient way to identify optimized transceiver locations as well as suitable radio parameters for C-ITS networks. At the same time, the simulation can be parameterized and reproduced as a planning tool for different scenarios and propagation models. Conventionally, radio planning is applied in the field of mobile radio for the determination of base station locations for mobile radio cells. Requirements are set here with respect to radio coverage and capacity utilization [1, 2, 3]. Further applications include indoor radio propagation, e.g. for localization systems [4, 5]. On the other hand, the results of the radio planning for C-ITS sites performed in this paper are evaluated for communication services based on the ETSI ITS-G5 standard developed for Vehicle Ad-hoc-networks (VANET) in the 5.9 GHz frequency range [6]. Prior work in die field of predicting reception qualities for vehicular communications has been done in [7] by collecting ITS-G5 radio measurements and developing a measurement based radio channel model. Simulation of signal reception quality is performed by using machine learning process. Research by [8] presents a simulation-based evaluation to measure the impact of geometry-based propagation model on connected vehicle performance in an urban environment. Communication quality is analyzed in terms of communications and networking performance. However, the presented approach does not consider measurement-based evaluation of the performed radio simulation. In [9], simulations and comparative measurements of radio propagation for urban C2X links were performed. The results are based on a simulation framework for ITS-G5 with given signal propagation models and are represented by connectivity maps.

Literature review has shown that simulations of radio links are often used to evaluate technology-dependent protocol parameters. This is based on an accurate radio planning simulation. As an improvement of existing work, this paper describes a calibration approach, which is usually insufficiently considered, to achieve an optimized realistic radio propagation simulation for C-ITS services. Since radio planning is used to assess the quality of service coverage, specific calibration characteristics of the radio modules used must be recorded to evaluate and optimize the simulation results if needed. These recordings are determined by means of a wired laboratory setup and considered as calibration offset in the evaluation of the radio propagation simulation.

Summarizing, the main contributions of this paper are:

- Coverage-optimized C-ITS specific radio propagation and network modeling as a baseline tool for facility planning and equipping.

- Effective Calibration of both simulated and measured service received signal strengths based on a laboratory measurement setup and a self-surveyed CW measurement along a 3.5 km long test field for automated and connected driving in rural environments.

The remainder of the paper is structured as follows: After the introduction (sec. 1), necessary fundamentals on radio planning and modeling, including simulation and measurement calibration method, are presented in sec. 2. Subsequently, sec. 3 introduces implementational details for radio planning as well as the utilized calibration, CW and services measurement setup. In sec. 4 the quantitative results of the propagation simulation and the resulting calibration are presented. The paper concludes with a summary and proposals for future work in sec. 5.

**2. Fundamentals**

*Radio Planning and Modeling*

Investigating radio propagation effects and network planning can be realized by utilizing radio propagation simulations, which can generally be classified as numerical approaches aiming to directly solve Maxwell equations or empirical approaches relying on surveyed measurements. While numerical approaches are computationally heavy and potentially cannot be solved in complex real-world scenarios, empirical and semi-empirical approaches heavily rely on the utilized hardware and propagation conditions of the surveyed data [10, 11]. In addition, ray-tracing and stochastic approaches with respect to the propagation environments can be applied [12]. These approaches allow the consideration of different hardware configurations and propagation environments and therefore provide a reasonable trade-off between flexibility, complexity and accuracy.

The former approaches are realized by radio propagation simulation tools, which perform propagation simulations based on an environmental and a radio model. As a first step, a three-dimensional CAD replica of the on-site conditions within the planning area has to be created. 3D-Objects relevant to radio network planning are identified and digitally modelled. In order to consider the influences of the topographical conditions, a digital terrain model (DTM) can be included. Particularly in urban environments or other areas with frequent and abrupt height differences, a high resolution DTM usage is suggested. Even in rural areas, typical landscape topographies such as hills or mountain ranges can cause radio shadowing, which cannot be mapped with sufficient accuracy without a DTM.

Next to the geometric dimensions, electro-magnetic properties for each object respectively all parts of the objects need to be assigned. For this, radio propagation simulation tools generally provide material catalog, which includes the electrical permittivity and permeability of the medium. If proper values are not available, these parameters can be determined by using permittivity measurements.

In addition, the simulated radio properties describe the physical characteristics of the radio technology under study, which includes frequency band, transmission power and antenna type, pattern and specifications, e.g. directional antenna.

Alongside the radio properties, the type of simulation and simulation parameter influence the conformity of the simulation and the real-world propagation as a trade-off with computational complexity. In this work, the Dominant Path Model (DPM) is used, which is a parameterizable, accurate and fast radio propagation model for calculating the direct signal paths between transmitter and receiver with regards to the environmental model. The main features of the model are an unlimited number of interactions (changes of orientation) along the path, a consideration of optimal wave guidance (wave guidance factor, based on reflection losses of the elements) as well as robustness to errors of the vector database used.

When comparing radio propagation models, preliminary research in [13] has shown that a DPM tends to produce more pessimistic simulation results compared to a simplified 3D ray-tracing due to the non-observance of multipath effects. However, these conservative estimates allow better conclusions regarding the minimum signal reception quality, while the ray-tracing simulation usually overestimated the actual received signal levels. Furthermore, according to [14] the lower computational costs with short prediction times and high accuracy favor the application of an urban DPM over ray-tracing, which is preferably used for multipath propagations (reflections, diffractions, scattering). An acceleration of the computation time can be achieved with an intelligent ray-tracing model. However, this model shows a high accuracy only in the range of the transmitter. With increasing distance, this accuracy is limited [14]. Investigations of the applicability of a Two-Ray Path Loss Model were considered in [15]. It was found that the most commonly used simplified Two-Ray Ground Model has no additional value over the much easier Free-space model. However, an extended Two-Ray model can be advantageous in urban environments where multipath effects are more frequent.

*Path Loss Modeling*

To evaluate the investigations and measurements made in this work, a view at theoretical and empirical propagation models is required. A baseline for modelling the received signal strength (RSS) is provided by the link budget, which calculates the received power $P_R$ as the sum of the transmitted power $P_T$, the cable and connector losses on the transmitter and receiver side $C_T$, $C_R$, the antenna gains $G_T$, $G_R$ and the path losses due to free-space attenuation $PL_{FS}$ and other combined signal path losses $PL_{Div}$ [16]:

$$P_R = P_T - C_T + G_T - PL_{FS} - PL_{Div} + G_R - C_R . \qquad (1)$$

The frequency (*f*)- and distance (*d*)-dependant empirical modeling of the path loss *PL* [dB], assuming free space propagation is given as follows [17]:

$$PL_{FS} [dB] = 20 \lg (d\,[m]) + 20 \lg (f\,[Hz]) - 147.55 \text{ dB}, \qquad (2)$$

where -147.55 dB is calculated with the speed of light $c_0$ [m/s] from $20 \lg (4\pi/c_0)$.

In real measurement scenarios, other attenuation factors $L_{Div}$ such as multipath effects, signal shadowing and other signal propagation phenomena can occur, most of which are also conditional on the propagation environment (indoor, urban, suburban, rural). These effects can partially be modeled as additional path losses using deterministic signal propagation models.

According to [18], the path loss *PL* of DPM, which is used in this paper, is calculated as:

$$PL\ [dB] = 20\ lg\ (f\ [Hz]) - 147.55\ \text{dB} + 10\ p\ lg\ (d\ [m]) + \sum_{i=1}^{n} g(\Delta\varphi, i)\ [dB] + \Omega\ [dB], \quad (3)$$

where *d* is the path length of a signal path. An individual path loss exponent *p* for line-of-sight (LOS) and non-line-of-sight (NLOS) conditions, before and after a certain breakpoint, has to be defined. Individual interaction losses for each interaction *i* of all *n* interactions are given with the function $g(\Delta\varphi, i)$, where $\Delta\varphi$ is the angular difference before and after the interaction. $\Omega$ is the gain due to waveguiding, considering different rays for each dominant path guided by reflections at the walls. The waveguiding depends on their material (reflection loss), their orientation (reflection angle) and the distance between the walls and the path.

*Simulation Path Loss Calibration*

In addition to the theoretical and simulative examination of RSS, the transmission of CW signals at a given radio frequency offers the possibility to measure real-world RSS in application-oriented transmission environments. Thus, location-based CW measurement values are available, each of which is the result of the power balance equation and thus contains the real path loss.

By comparing CW measured values with a radio propagation simulation with predefined model parameters carried out in advance, model calibration can be performed by adjusting parameters *p* and $g(\Delta\varphi, i)$ in the DPM path loss model in order to better map measured value distribution. Since model parametrization on its own is not able to fully represent CW measured values and simulation results are often conservative estimations, an additional calibration step is suggested, which is performed by determining a constant path-loss offset with percentage weighting factor $W \in 0,\ldots,1$. The offset is calculated by comparing the simulation path loss $PL_{sim}$ and the measured CW path losses $PL_{CW}$ given the position *k*. This allows, for example, along a trajectory ($k \in 1,\ldots,K$) the determination of an average weighted difference

$$\Delta\overline{PL}_{cal} = \frac{W}{K}\sum_{k=1}^{K}\left(PL_{CW,k} - PL_{sim,k}\right), \quad (4)$$

which can be taken as additional calibration value of the simulated received levels $P_{R,sim}$. Accordingly,

$$P_{R,sim,cal,k} = P_{R,sim,k} + \Delta\overline{PL}_{cal} \quad (5)$$

is the calibrated signal reception level of the radio propagation simulation in dBm.

This approach of calibrated radio planning simulation can increase the accuracy of simulation results and optimize planning for C-ITS infrastructure components at different sites. However, a disadvantage of this method is that the average calibration value obtained does not provide a universally optimal representation of the real CW measured values for each simulation location. Moreover, the calibration value is also calculated in NLOS conditions where $\Delta\overline{PL}_{cal}$ is usually larger. Therefore, it is reasonable to determine different average calibration values depending on visibility conditions and use them for simulation calibration. In addition, a larger number of CW measurement points can increase the statistical confidence for determining the calibration values. The currently available radio network planning tools are occasionally able to calculate and consider correction factors and parameters independently, based on the processing of real CW measured values.

*Service Path Loss Calibration*

The performance of a radio propagation simulation as the basis for site planning of ITS infrastructure components focuses especially on the optimal coverage of communication services. In the field of wireless C-ITS communication, services in the frequency range of 5.9 GHz are considered in this paper. Radio modules supporting these services are installed as C-ITS components at the RSU locations identified by radio planning. Accordingly, it is necessary to evaluate the radio performance on both the transmitting and receiving sides in terms of RSS. This allows a comparison of radio planning results and signal levels provided by the radio modules.

Radio modules output a received signal strength indicator (RSSI) as a level measurement value. However, this often does not correspond to the RF level measurement values determined by RF measuring equipment. Therefore, it is also necessary to perform a calibration measurement. In a wired measurement setup, both the transmit and receive power $P_{R,module}$ specified by a radio module are verified with the RF receive power of a parallel-connected spectrum analyzer $P_{R,spec}$ for different levels of simulated signal attenuation $m \in 1, ..., M$. The difference of the determined received values describes the calibration offset:

$$\Delta \overline{P}_{R,cal,module} = \frac{1}{M} \sum_{m=1}^{M} \left( P_{R,spec} - P_{R,module} \right). \tag{6}$$

The calibrated received power of a radio propagation simulation for C-ITS modules is thus given by

$$P_{R,sim,cal,module,k} = P_{R,sim,cal,k} + \Delta \overline{P}_{R,cal,module}. \tag{7}$$

Radio modules have a characteristic receiver sensitivity $P_{R,module,sens}$, which depends on the modulation and coding scheme (MCS) applied. It specifies an RSSI below which the packet error rate increases significantly. When selecting locations for RSU, the following condition should accordingly be met at least at each position $k$ in order to be able to guarantee sufficient radio quality:

$$P_{R,sim,cal,module,k} > P_{R,module,sens} + \Delta \overline{P}_{R,cal,module}. \tag{8}$$

## 3. Implementation and Measurement setup

The test environment extends in a real C-ITS corridor along a national road (Bundesstrasse B170) in the south of the city of Dresden, Germany. The section of road modelled within the radio network planning has a total length of 3.5 km and contains a large number of different traffic elements, such as crash and noise barriers, vegetation, cars, lorries, busses, traffic lights, traffic signs, buildings, bridges and many more. The area of study was modelled realistically with the help of 3D CAD objects and parameterized with the help of various object properties.

For the radio propagation simulation performed in this work, the software WinProp is used [19]. In order to represent the most detailed possible representation of road traffic and environmental elements for the 3D model, an urban configuration database was used in WinProp, which, in addition to the DTM, can also take very specific 3D objects and better NLOS propagation characteristics into account. With WinProp, specific permittivity properties ε can be assigned to objects or parts of them via substance assignments. In addition to individual object properties, the tool also offers a variety of parameterization options for transmitters and antennas. Modeling of the simulation results considered in this paper is performed by using transmitter-side radio planning parameters listed in Tab. 1. As introduced beforehand, urban DPM was used as the propagation model. The following path loss exponents $p$ were used within the simulations, where obstructed-line-of-sight (OLOS) describes transmitter-receiver connection in the same corridor without wall intersection and without having direct LOS:

- Default values $p$ as recommended in [20] for low transmitter: LOS 2.6; OLOS 2.8;
- Calibrated values $p$: LOS 2.3; OLOS 2.9.

*Simulation and field test environment:*

A three-armed, signalized intersection area within the modeled corridor was chosen for the measurement due to its topological and radio propagation conditions. Many challenges and characteristics are met in the area of study that exist on the roadway side, including a potential backlog of vehicles at the traffic signal stop lines and noise barriers as protective elements along the left lane. In addition, the intersection area was selected as a potential RSU location. Thus, both LOS and NLOS conditions can be observed along the road. The 3D-intersection model and uncalibrated simulation is shown in Fig. 1.

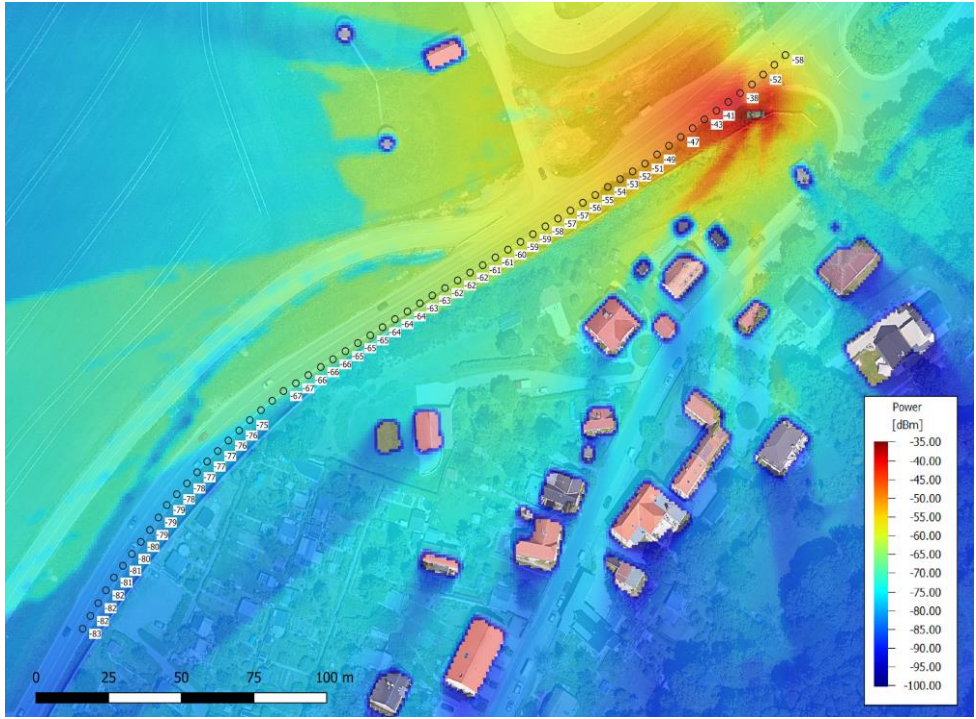

**Figure 1 - Radio propagation simulation of the intersection area with default parametrization.**

On the transmitting side, a directional transmitting antenna was mounted next to the traffic signal on a tripod. The antenna was oriented to the south along the traffic signal. The alignment of the antenna to the south along the road also takes into account the slight gradient of the roadway. For the CW measurements, a CW signal generator [21] with transmitting power $P_T$ =10 dBm was connected to the transmitting antenna. Alternatively, a RSU [22] was connected to the antenna for ETSI ITS-G5 service measurements and transmitted Cooperative Awareness Messages (CAM) with a frequency of 10 Hz with $P_T$ = 23 dBm. MCS QPSK with coding rate (CR) = ½ was used, with $P_{R,module,sens}$ = -95 dBm for one antenna in LOS condition. Furthermore, positions were recorded with the internal GPS receiver.

**Table 1 - Configuration parameters of the simulation and field test measurements.**

| general parameters | | transmitting site | receiving site |
|---|---|---|---|
| antenna | height | 4 m | 1.5 m |
| | type | panel (directional) | Dipole |
| | gain $G$ | 10 dBi, | 2 dBi |
| | directivity | 16° horizontal/vertical directional beam | omnidirectional |
| | polarization | vertical | vertical |
| transmitting power $P_T$ | | 23 dBm | - |
| signal frequency $f$ | | 5.9 GHz | |

On the receiver side, a mobile research vehicle was converted for measurement purposes, with an omnidirectional dipole antenna mounted on the vehicle roof. A spectrum analyzer [23] located in the vehicle measures the received CW signal level. Measurements have been made at equidistant intervals of 5 m along the roadway. The C-ITS service measurement was performed in an independent measurement run using a OBU [24] as the receiving unit. Signal reception levels were recorded as module-specific RSSI for each received message in addition to GPS positions of the vehicle.

The measurement setup of the field test measurements is shown in Fig. 2. Relevant configuration parameters for the transmitter and receiver antennas can be seen in Tab. 1 for both CW and service measurements. Both measurements were recorded under real conditions in flowing traffic.

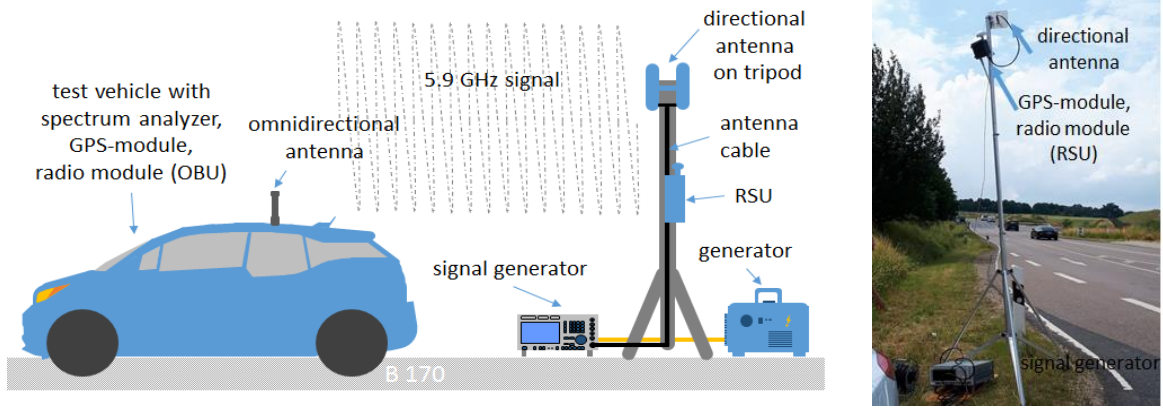

**Figure 2 - Field test measurement setup.**

*Calibration Measurements of the Radio Modules*

Validation of the accuracy of the receive performance of the C-ITS radio module was realized by a calibration setup in a laboratory environment. The objective is to determine a calibration offset $\overline{P}_{R,cal}$. Fig. 3 shows the setup of the measurement instruments used. Here, a OBU, configured as a transmitter is connected to a controllable step attenuator [25] to simulate the attenuations occurring in a wireless experimental setup. The 5.9 GHz signal is routed both in the signal analyzer and wired to the second radio module configured as a receiver. During the measurement, the transmitter OBU was configured to send 100 message packets with payload size 500 bytes each with a message frequency of 10 Hz, a transmission power $P_T$ = 23 dBm and MCS QPSK, CR = 1/2. Signal attenuation $m$ was set in steps of 1 dB in the range $m \in 35, \ldots, 105$ dB. The evaluation is performed on the receiver side in the radio module by averaging the RSSI of all received message packets. The spectrum analyzer records the maximum level during the entire transmission process. Signal attenuation due to the line-conducted setup has been taken into account. The calibration setup further allows validation of the signal output power of the transmit module by measuring the signal input power at the spectrum analyzer.

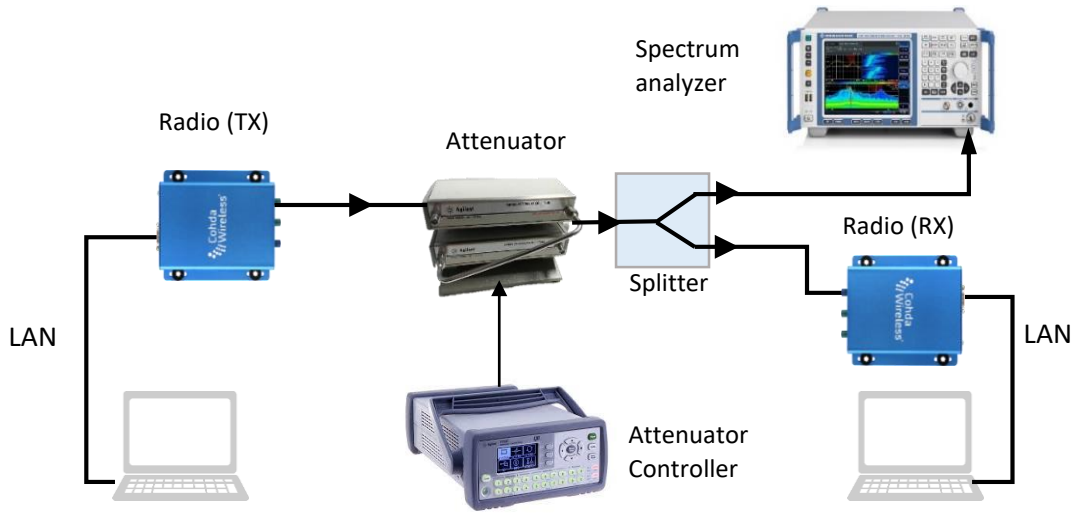

**Figure 3 - Measurement setup for calibration measurements of the radio modules.**

## 4. Investigation Results

*Modeling and Simulation*

The simulated RSS for the default set of parameters of the DPM, given in sec. 3, are plotted in Fig. 1. Here, along the roadway, the simulated received levels at the measurement points of the subsequently performed CW measurement are indicated. The simulation shows a sufficiently good reception quality up to $P_{R,sim}$ = -95 dBm in a distance of about $d$ = 245m from the transmitting antenna. Due to the noise barrier along the curve area NLOS conditions occur, where the simulated signal levels drop significantly. The CW measurements necessary for the calibration of the radio propagation model were carried out along the road. Fig. 4 (left) shows the measured RSS levels at defined measurement points in the test environment. It should be noted that the transmit power of the signal generator was mathematically adjusted to the higher transmit power of the radio modules for comparability of the measured values. We observed RSS values $P_{R,spec}$ from -35 to -93 dBm during the measurement run.

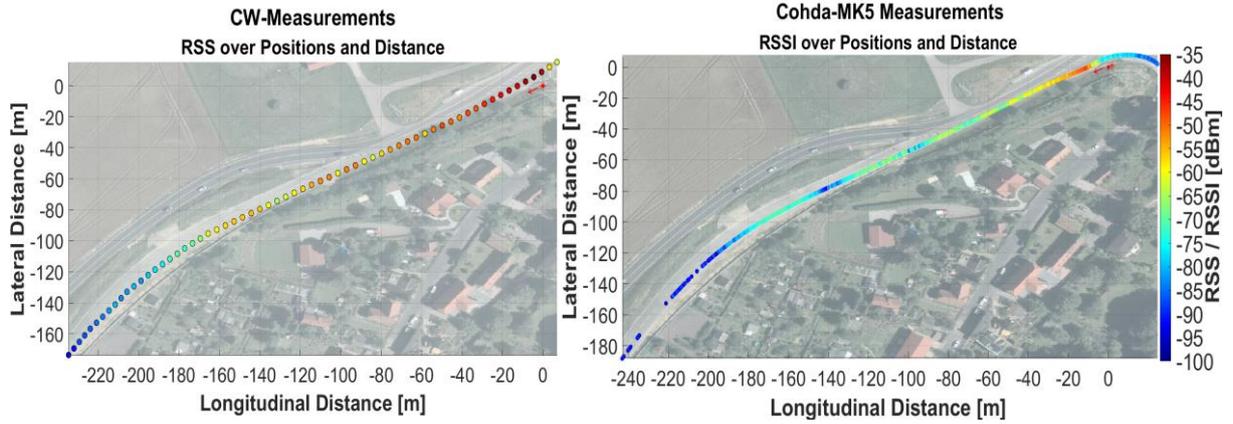

**Figure 4 - CW (left) and ITS-G5 service (right) measurements in the testbed.**

Comparing the simulated path loss (standard model parametrization) with the CW measured path loss, a relatively conservative simulation can be observed for the path loss in Fig. 7. In LOS conditions the measured values are mainly above the simulation. However, for NLOS this relationship is just the exact opposite. The corresponding root-mean-square error (RMSE) and standard deviation (SD) are given in table 2 for the respective model parametrization. Therefore, the calibration of the radio propagation model is necessary. The goal is to adjust the simulation parameters to approximately fit the measured path losses. Parametrization of the path loss exponents for LOS, OLOS and NLOS conditions by the parameters given in sec. 3 leads to an improvement of the ratio between simulated path losses versus measured values. Especially in the NLOS condition, the calibrated propagation model shows improved performance. Additionally, the mathematically determined calibration offset was taken into account, which resulted in a mean weighted difference of $\Delta\overline{PL}_{cal}$ = 0.5 dB and was attributed to the parametrized DPM with a weighting factor of $W$ = 0.25 for a conservative estimate. It should be noted that these calculated difference is small due to the low weighting and can thus be considered negligible in practical use. However, conservative estimation of path losses relative to actual measurements is beneficial for the application of radio planning as a tool for determining C-ITS locations to ensure safety margins in radio coverage. The resulting calibrated path losses of the simulation are plotted as a red dashed line in Fig. 7 for the corresponding measurement distances with respect to the RSU.

*Calibration Measurement of Radio Modules*

The result of the calibration measurement shows that the output received power of the radio module continuously falls below the configured transmit power in the wired configuration. The attenuation losses due to cables and RF components are already taken into account. As can be seen in Fig. 5 the difference to the received power of the spectrum analyzer averages around $\Delta\overline{P}_{R,cal,module}$ = 8 dBm. Here, the difference remains largely constant. Only at the receive sensitivity limit, small deviations can be observed. The received signal level of the spectrum analyzer is on average 1 dB above the theoretical receive level $P_R$ according to (1) with known transmit power of the radio module and link losses.
The dimensioning of C-ITS networks for communication services in the 5.9 GHz ETSI ITS-G5 standard with the aid of a radio planning simulation therefore requires the consideration of the calibrated radio module RSSI. For the radio planning of RSU sites with the radio modules used, the CW calibrated receive levels must therefore be corrected downward by $\Delta\overline{P}_{R,cal}$ in order to correspond to the module calculated RSSI value and to be able to map their critical receive sensitivity. For the OBU and MCS used, this is specified as $P_{R,module,sens}$ = -95 dBm. Below this level packet error rate exceeds a value of 10 % and increases significantly [22].

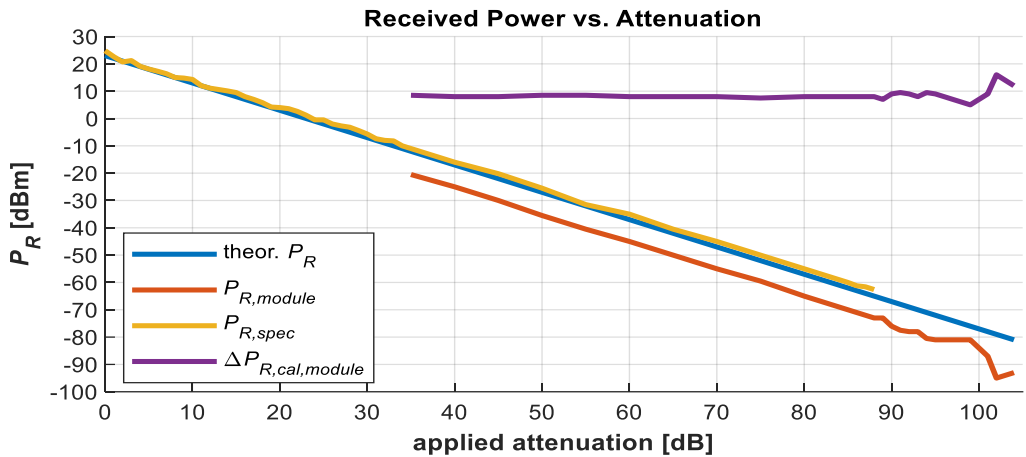

**Figure 5 - Module calibration measurements.**

*Service coverage simulation and measurement calibration*

In order to simulate RSSI of the MK5 radio modules for ETSI-ITS-G5 communications services, the module-specific calibration factor $\Delta \overline{P}_{R,cal,module}$ was taken into account in the CW-calibrated radio propagation simulation. The resulting simulation is shown in Fig. 6, where $P_{R,sim,cal,module}$ is indicated for positions of the performed service measurements. To validate the accuracy of the module calibrated simulation, the OBU measurement data recorded in the field test are compared. Fig. 4 (right) shows the RSSI measurements along the road. As expected, the signal level decreases with increasing distance to the RSU, with significant signal attenuation in the area of the NLOS. The signal power reaches the receiver sensitivity limit $P_{R,module,sens}$ at a distance of 220 m in lateral direction, which implies that measurement data acquisition does not occur here continuously as the distance increases. RSSI $P_{R,module}$ from -45 to -96 dBm were observed during the measurement run.

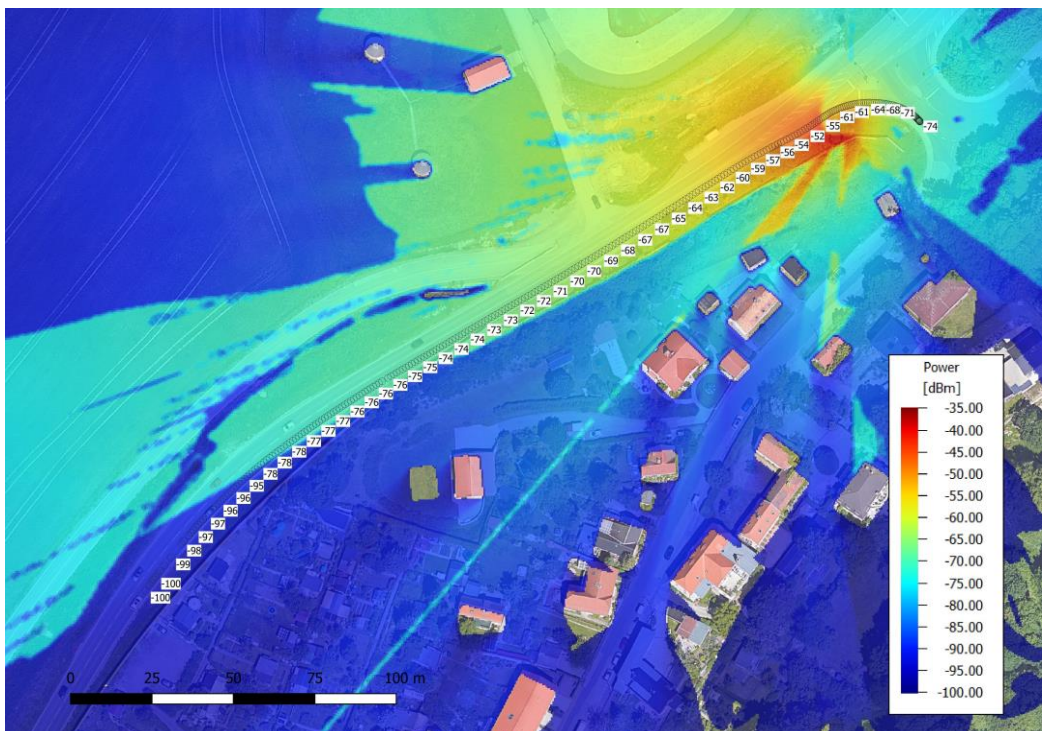

**Figure 6 - Calibrated radio propagation simulation for ITS-G5 radio modules.**

By comparing simulated and measured values in terms of path loss, calibrated simulation fits the measurements as indicated in Tab. 2. Path loss values are plotted in Fig. 7, where we can observe large variability in the measured values. However, the module calibrated DPM path loss with additional constant offset roughly fit a logarithmic approximation of the measured value variances over the distance for LOS conditions. For NLOS, the model overestimates the actual measured *PL* values. In these areas, however, a safety margin is advantageous for radio planning purposes in order to avoid regions with a high probability of signal interruptions when planning C-ITS networks.

Table 2 - Evaluation measures of the simulation model for respective visibility conditions.

|  | standard simulation vs. CW measurement | | | calibrated simulation vs. CW-measurement | | | module calibrated simulation vs. OBU measurement | | |
|---|---|---|---|---|---|---|---|---|---|
|  | all | LOS | NLOS | all | LOS | NLOS | all | LOS | NLOS |
| **RMSE [dB]** | 4.5 | 3.9 | 5.7 | 3.2 | 3.1 | 3.2 | 4.4 | 3.8 | 6.5 |
| **SD [dB]** | 5.6 | 3.7 | 4.4 | 3.8 | 3.2 | 3.7 | 5.0 | 4.6 | 4.1 |

The comparison of the CW and RSSI measurement shows an average deviation of the path loss levels over all distances of $\Delta PL = 9$ dB. Thus, the calibration value $\Delta \overline{P}_{R,cal}$ determined under laboratory conditions could be approximately determined under real conditions as well. The deviation of the two values can be assumed in the two-stage test execution with slightly different measurement positions and intervals of the CW and service measurements, respectively, as well as the different traffic conditions. Furthermore, additional attenuations, e.g. due to occurring multipath effects or unmodeled vegetation and vehicles along the roadway, can be observed at the same position in both measurement series.

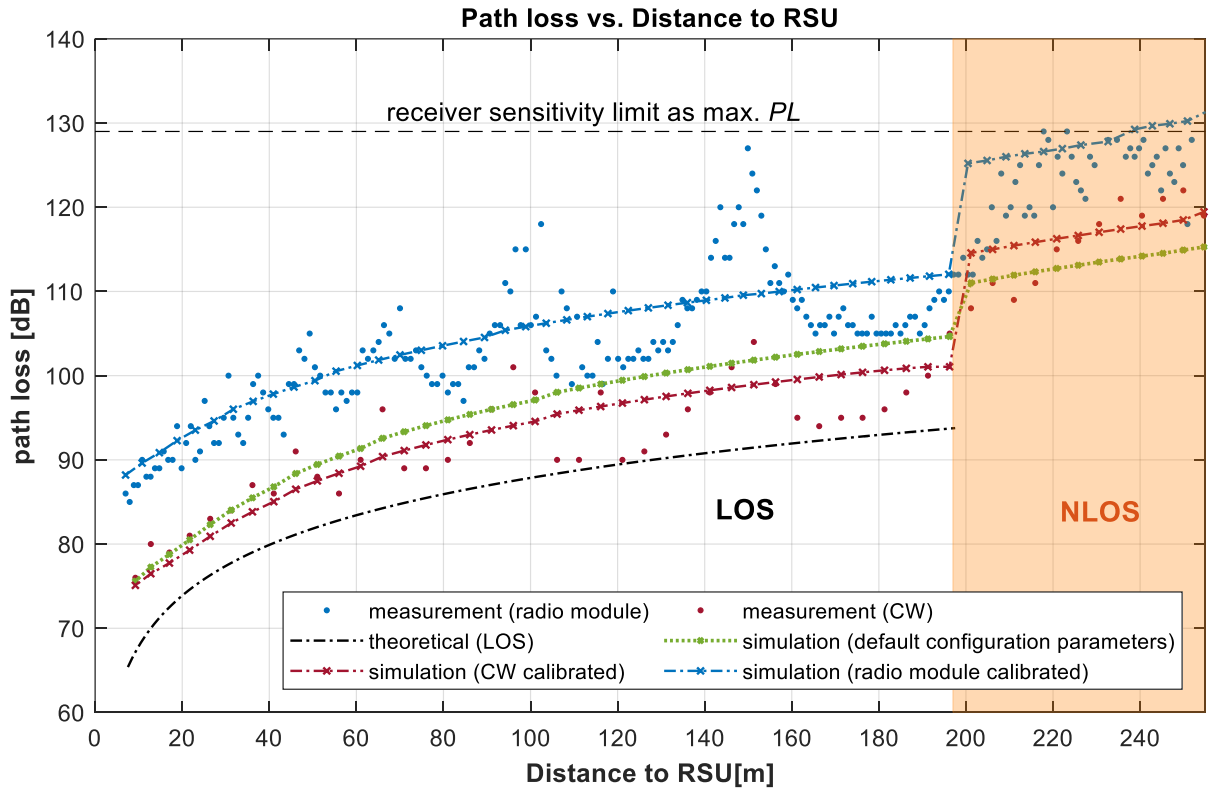

Figure 7 - Path loss over distance for simulations and measurements.

To plan RSU locations for C-ITS applications, areas of critical radio undersupply must be identified by simulation. This is necessary for all considered subsections of the planning corridor. Therefore, the receiver sensitivity is a module and MCS dependent evaluation parameter. In the module calibrated radio propagation simulation, this limit is exceeded along the road at a distance of approx. 240 m relative to RSU position with corresponding $PL$ = 129 dB. To ensure complete radio coverage along the entire road, another RSU at the next intersection should provide radio coverage for this area. However, the use of uncalibrated models calculates a lower path loss in NLOS condition, so that too good radio coverage is simulated for the radio modules used. In a real scenario, this leads to an undersupply of certain areas.

**5. Conclusion and Outlook**

The methods presented in this paper for calibrating software-based radio propagation simulations provide an efficient way to plan RSU locations for radio modules that enable communication services according to the ETSI ITS-G5 standard in the 5.9 GHz frequency band. First, the basis of the investigations is an accurate model of the simulation environment with all relevant objects and their electrical permittivity. In the second step, it is necessary to calibrate the signal propagation model of the radio simulation. This is done by comparing with recorded CW measured values. Thus model parameters can be adjusted and additional calibration offsets can be determined. Finally, the calibrated models are applied to communication services simulation. Since the used radio modules report a RSSI with an offset-level, a second calibration step is necessary. Therefore, RSS-offset has to be determined in a laboratory measurement setup. As a result of the simulation and performed calibration, it is possible to determine further RSU locations in C-ITS networks for intelligent infrastructures efficiently and without metrological effort. Thus, radio propagation planning can provide great value as a powerful engineering tool and costly RF measurements in real environments can thus be mostly substituted. Nevertheless, it is reasonable to validate the determined calibration values with a larger measurement database and, if necessary, to calculate them again. Here it is also necessary to perform a module-specific calibration once. Under certain circumstances, it also seems reasonable to show the CW calibration values as a function of the ambient conditions.

Furthermore, other signal propagation models such as ray-tracing should be considered in future studies. To provide advantage for radio coverage flexibility, beam-switching and -steering antenna technology should also be addressed in future research. Alternatively, road sections with potentially weak radio coverage, could be supplied with leaky coaxial cables (LCX) as an alternative to conventional antenna technology to ensure scalable radio coverage in line with demand [26]. With the methods presented, other communication services besides ETSI ITS-G5 can be considered for a calibrated radio planning simulation, e.g. cellular mobile communications, especially 5G-networks. Above all, planning of base station locations for mobile radio units in the context of C-ITS is conceivable.